\RequirePackage[l2tabu, orthodox]{nag}
\documentclass{stsci_report}
\usepackage[pdftex,
pdfauthor={A. Guzman},
pdftitle={Measuring the Column Dependence of Read Noise in ACS/WFC Bias Frames},
pdfsubject={The noise in bias frames for all four readout amplifiers in the Advanced Camera for Surveys (ACS) Wide Field Channel (WFC) is dependent on row number. This is because dark current accumulated during readout increases across the detector, influencing and increasing the read noise as a function of row number. In this report, we investigate bias frames taken with the ACS/WFC to explore the column dependence of read noise for each of the amplifiers for different anneal periods. Analyzing the data we find that there is no column dependence of read noise and that the read noise values for the physical pre-scans is approximately 0.5 e$^-$ lower than in the science arrays because there is no readout dark accumulated in this area. We further investigate 1) the evolution of read noise over an anneal period, 2) a linear decrease in read noise within the initial columns per amplifier, and 3) pixels in elevated read noise columns. We conclude that 1) there is no visual trend of read noise over an anneal period, 2) amplifiers A and C have an initial linear decrease of read noise in the science arrays, and 3) masking unstable hot pixels in a column will decrease its read noise values.},
pdfkeywords={ACS, CCD, Hubble Space Telescope, HST, Space Telescope Science Institute, STScI, Advanced Camera for Surveys, Wide Field Channel, Read Noise, columns, WFC, readout, amplifiers}]{hyperref}
\usepackage[section]{placeins}
\usepackage{booktabs}
\usepackage{rotating}
\usepackage{natbib}
\usepackage{url}
\usepackage{graphicx}
\usepackage[font=footnotesize,labelfont=bf]{caption}
\usepackage{subcaption}
\usepackage{nameref}
\usepackage{hyperref} 
\usepackage{array} 
\usepackage{color}
\usepackage{amsmath}
\usepackage{gensymb}
\usepackage[bottom,symbol]{footmisc}

\usepackage{float}
\usepackage{aliascnt}
\newaliascnt{eqfloat}{equation}
\newfloat{eqfloat}{h}{eqflts}
\floatname{eqfloat}{Equation}
\usepackage{adjustbox}
\usepackage{threeparttable}
\usepackage{booktabs}

\copyrighttext{Copyright \copyright \the\year\ The Association of Universities for Research in Astronomy, Inc. All Rights Reserved.}

\title{\textbf{Measuring the Column Dependence of Read Noise in ACS/WFC Bias Frames}}
\presubtitle{\flushleft{\includegraphics[width=5cm]{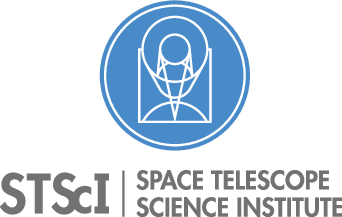}} \\ \hfill Instrument Science Report ACS 2023-03}
\author{A. M. Guzman, M. C. McDonald}
\date{June 22, 2023}

\begin{document}

\maketitle

\abstract{The noise in bias frames for all four readout amplifiers in the Advanced Camera for Surveys (ACS) Wide Field Channel (WFC) is dependent on row number. This is because dark current accumulated during readout increases across the detector, influencing and increasing the read noise as a function of row number. In this report, we investigate bias frames taken with the ACS/WFC to explore the column dependence of read noise for each of the amplifiers for different anneal periods. Analyzing the data, we find that there is no column dependence of read noise and that the read noise values for the physical pre-scans are approximately 0.5 e$^-$ lower than in the science arrays because there is no readout dark accumulated in this area. We further investigate 1) the evolution of read noise over an anneal period, 2) a linear decrease in read noise within the initial columns per amplifier, and 3) pixels in elevated read noise columns. We conclude that 1) there is no visual trend of read noise over an anneal period, 2) amplifiers A and C have an initial linear decrease of read noise in the science arrays, and 3) masking unstable hot pixels in a column will decrease its read noise values.}

\section{Introduction}
Since the installation of the Advanced Camera for Surveys Wide Field Channel (ACS/WFC) on the Hubble Space Telescope (HST), it has been operating in a Low Earth Orbit radiation environment where energetic particles in the radiation belts damage the detector. This radiation damage causes a linear increase of dark current over time, contributing to an increase in readout dark due to more warm and hot pixels being produced throughout the lifetime of ACS \citep{isr_jenna}.

\begin{figure}[H]
  	\centering
 	\begin{subfigure}{1\textwidth}
  		\centering
		\includegraphics[scale=0.10]{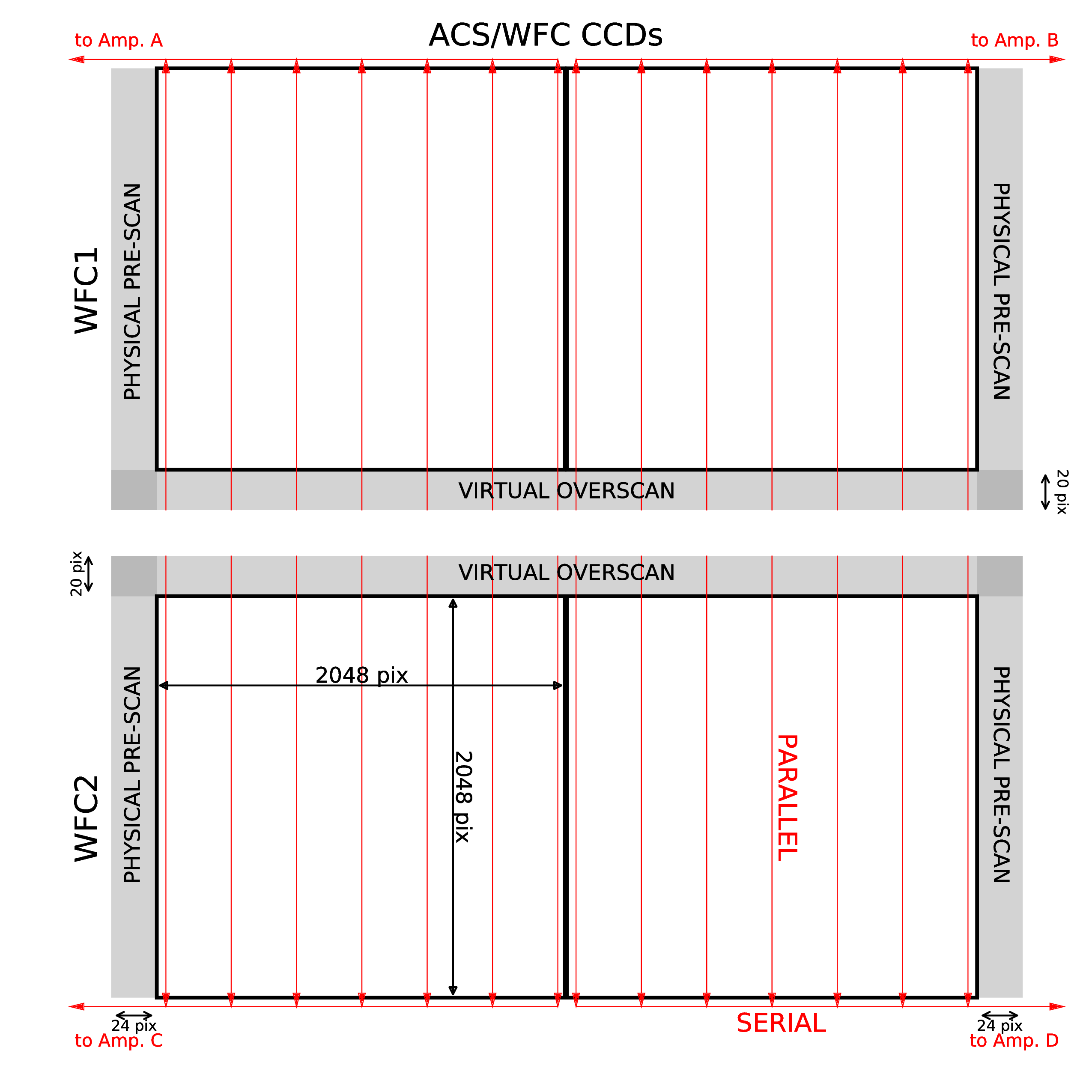}
		\end{subfigure}
 \caption{The layout of WFC full-frame images, including parallel and serial readout directions, physical prescans, and virtual overscans \citep{acsihb}.}
    \label{fig:acsccd}
\end{figure}

The WFC has two, 4144 x 2068 pix$^2$, charged-coupled devices (CCDs), WFC1 and WFC2. These CCDs have 24 physical pre-scan columns and 20 virtual overscan rows, which leaves an area of 4096 x 2048 pix$^2$ for the science frames. Each CCD has two amplifiers that read out 2072 x 2068 pix$^2$ of data. Amplifiers A and B correspond to WFC1, while amplifiers C and D correspond to WFC2 (Figure \ref{fig:acsccd}). In June 2007, the WFC stopped working due to malfunctions, but was recovered during the HST Servicing Mission 4 (SM4) on 2009.

 As part of the CCD Daily Monitor Program, four long (1000.5 s) and two short (0.5079 s) dark frames are taken Monday, Wednesday, and Friday of every week. These dark frames are then combined to create a superdark image for each anneal period that is used to analyze dark current and hot and warm pixels. 
 Two bias frames are also taken the same days as the dark frames for the CCD Daily Monitor Program. These images are taken with zero seconds exposure time and are combined to create a superbias image. Bias frames have bias structure and readout dark present, which can then be used to calculate the read noise for each single image. 
 
Read noise is the noise associated with reading out each pixel on the detector. It takes approximately 100 seconds to read out a full-frame of ACS/WFC pixel data and, during this time, dark current accumulates in the pixels resulting in excess readout dark. This creates a positive linear relationship where readout dark increases across the detector, which means pixels furthest away from the amplifiers (last to be read out) are subject to the most readout dark. The readout dark contributes to the read noise in each amplifier and therefore read noise and row numbers share the same positive linear relationship. We use this relationship to calculate the read noise value with respect to amplifier. 

The ACS/WFC undergoes monthly anneals to mitigate the accumulation of dark current and warm and hot pixels. During the anneal, the WFC CCDs and thermal electric coolers are powered off while heaters are turned on to warm the CCDs from approximately -81\degree C to about 20\degree C \citep{isr_meaghan}. These anneals are performed to temporarily reduce the dark current and the population of warm and hot pixels. These cycles do not repair 100 percent of all hot and warm pixels in the CCDs, therefore there is a growing population of permanent hot pixels in the ACS/WFC. The read noise values of each amplifier are steady to 1\%, except on five different occasions during the ACS/WFC lifetime \citep{acsihb}. In each event, the anneal process has eventually stabilized the fluctuating read noise values.

\section{Data and Analysis} \label{s:data}

In order to investigate the relationship between read noise and column number over time, we retrieved the raw bias frames from four different anneal periods: a recent anneal period from when this project began (2022-08-17), an anneal period six months prior to that (2022-03-28), an anneal period from six years ago (2017-09-21), and an anneal period pre-SM4 (2005-04-19). To calculate the read noise, we created a difference image where we subtracted subsequent bias files over an anneal period. Then, we measured and subtracted bias striping while also mimicking the \texttt{acsrej}
step to identify and mask cosmic rays by sigma clipping the data. We looped through each column of the data and flipped the x-axis for amplifiers B and D since they are read out in the opposite direction (see Figure \ref{fig:acsccd}). Lastly, we took the variance of the data with respect to each column and ultimately used Equation \ref{eq:1} to calculate the read noise, where \emph{sciencedata} refers to the subtraction of two bias images. The variance is divided by two to account for the two bias images that were differenced.

\begin{eqfloat}
\begin{equation}\label{eq:1}
read noise (sciencedata) = \sqrt {\frac{variance(science data)}{2}}
\end{equation}
\end{eqfloat}

We analyzed the Data Quality (DQ) arrays of the superbias and superdark images because it is where the pixel flags are stored, and ultimately used to calibrate science images in \texttt{calacs}. Hot and warm pixels are flagged as 16 and 64 in the superdark DQ, respectively, and unstable hot columns are flagged as 128 in the superbias DQ arrays. Warm pixels contain a dark current range of $0.06-0.14 e^{-} /pixel/s$, and hot pixels have a higher dark current, $0.14 e^{-} /pixel/s$. 

\section{Initial Results}

\begin{figure}[!ht]
  	\centering
 	\begin{subfigure}{1\textwidth}
  		\centering
		\includegraphics[scale=0.43]{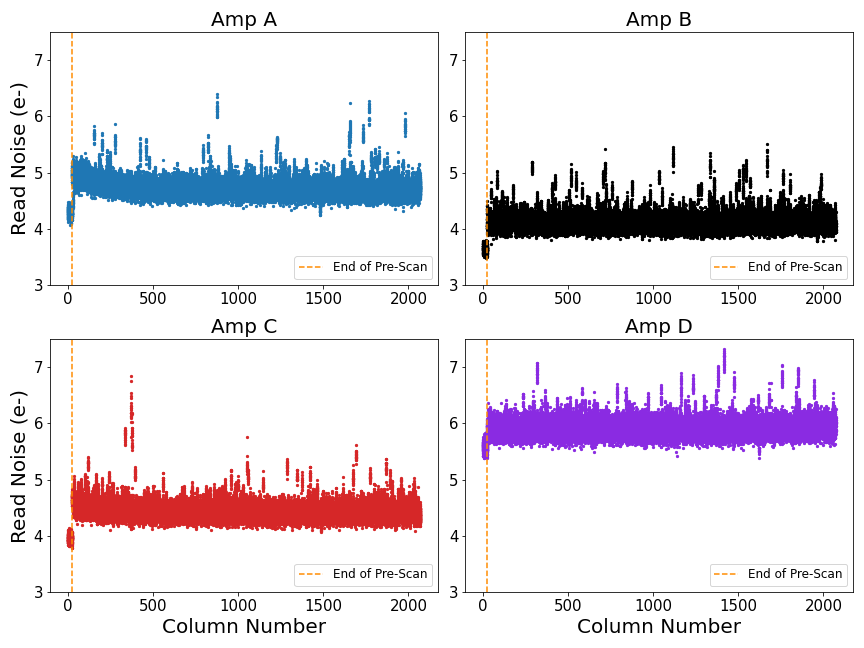}
		\end{subfigure}
 \caption{Data from the 2022-08-17 anneal period. Each point in a column refers to data from a difference image of two raw biases. Each plot with respect to amplifier was scaled the same to easily identify differences. The orange dashed-line separates the physical pre-scans columns (left) from the rest of the science columns (right).}
    \label{fig:202208anneal}
\end{figure}

We present the results of read noise as a function of column number for each amplifier for each of the four anneal periods in Figure \ref{fig:202208anneal}. The data indicates that for each amplifier, there is no significant column dependence of read noise to be seen. This was initially expected because the readout dark does not depend on column number, only on row number \citep{isr_jenna}. We observe that the physical pre-scans for each amplifier have a lower average read noise value than that of the science columns, and this is because physical pre-scans have no readout dark present. Due to this, we recommend not including the physical pre-scans when determining the read noise of each amplifier.  

\begin{figure}[ht]
  	\centering
 	\begin{subfigure}{1\textwidth}
  		\centering
		\includegraphics[scale=0.43]{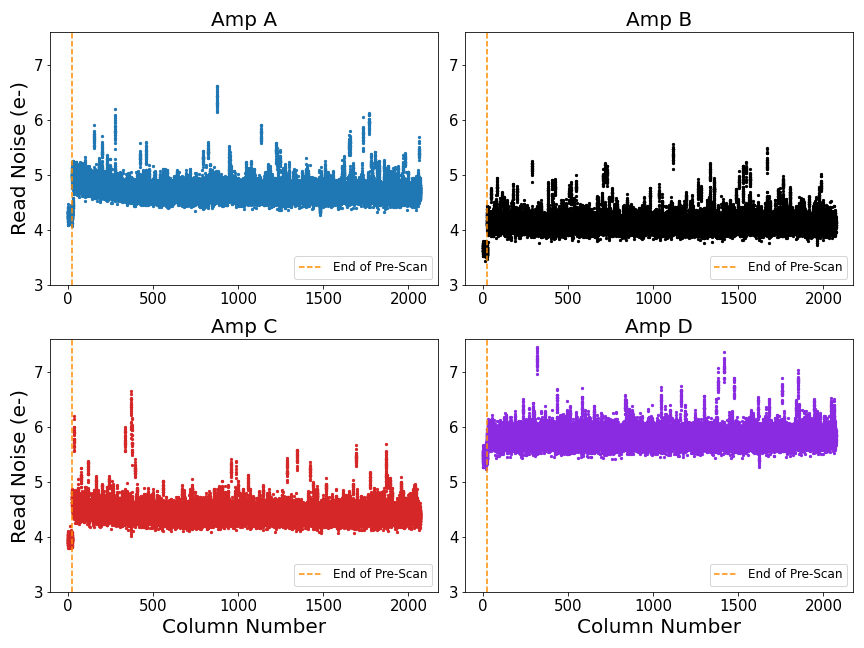}
		\end{subfigure}
    \caption{Data from the 2022-03-28 anneal period. All plot characteristics are the same as described in Figure \ref{fig:202208anneal}.}
    \label{fig:202203anneal}
\end{figure}

\begin{figure}[htb]
  	\centering
 	\begin{subfigure}{1\textwidth}
  		\centering
		\includegraphics[scale=0.43]{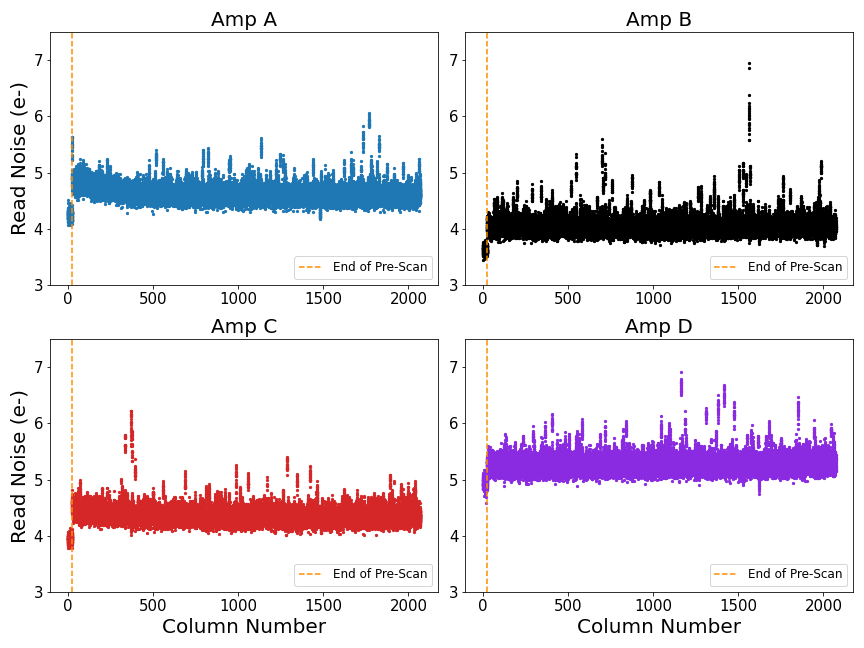}
		\end{subfigure}
    \caption{Data from the 2017-09-21 anneal period. All plot characteristics are the same as described in Figure \ref{fig:202208anneal}.}
    \label{fig:2017anneal}
\end{figure}
\begin{figure}[htb]
  	\centering
 	\begin{subfigure}{1\textwidth}
  		\centering
		\includegraphics[scale=0.43]{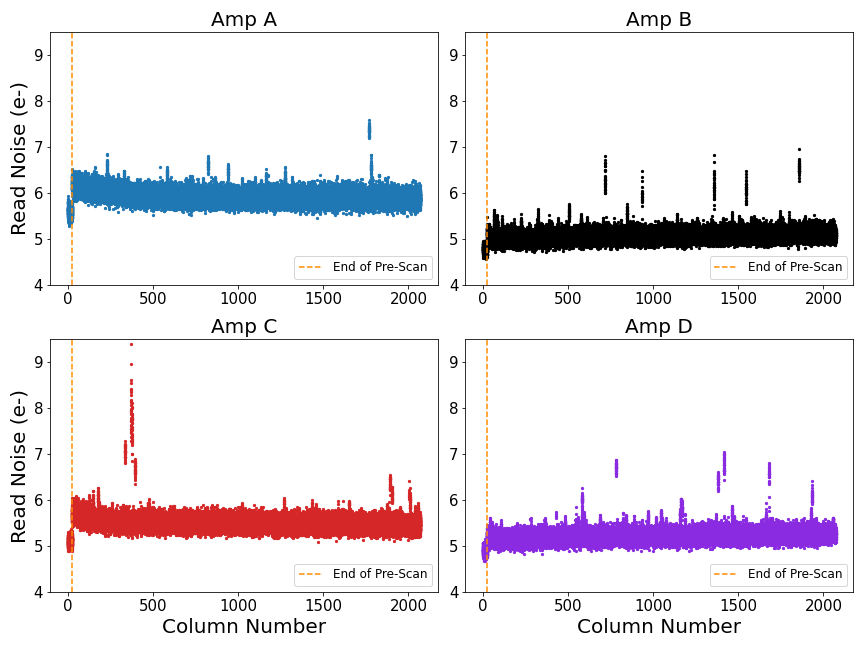}
		\end{subfigure}
    \caption{Data from the 2005-04-19 (Pre-SM4) anneal period. All plot characteristics are the same as described in Figure \ref{fig:202208anneal}.}
    \label{fig:2005anneal}
\end{figure}

We observe that specific columns had elevated read noise values that would change or remain the same throughout the anneal periods and decided to explore this further in the report. Amplifiers A and C appear to have an elevated initial dark rate in the first several hundred columns. This behavior is not observed in amplifiers B and D. Figures \ref{fig:202203anneal}, \ref{fig:2017anneal} and \ref{fig:2005anneal} are the read noise versus column number plots for the 2022-03-28, 2017-09-21 and 2005-04-19 anneal periods respectively. These plots share the same characteristics and trends as that of the 2022-08-17 anneal period.

\section{Further Analysis}

\subsection{Read Noise Evolution over an Anneal Period}

In addition to investigating the relationship between read noise and column number, we wanted to find out how the read noise in certain columns that have unique characteristics behaved. Characteristics like columns having elevated read noise on only one anneal, columns with elevated read noise in all anneals and columns with read noise that elevated overtime and continues to stay elevated. To do this, we organized all raw bias files in chronological order for each anneal period. For each amplifier, a specific column was chosen and its read noise values were calculated with respect to every bias frame taken during the anneal period. For amplifier A, we chose to work with column 873 because it was only elevated in the 2022 anneal periods. For amplifier B, column 1565 was investigated because it was only elevated in the 2017 anneal period; its read noise levels remaining average in the pre-SM4 and both 2022 anneal periods. For amplifier C we analyzed column 373 which was elevated in all the anneal periods. Finally, for amplifier D we focused on column 1425, which had average, non-elevated, read noise levels for all four anneal periods. 

\subsubsection{Results}
\begin{figure}[H]
  	\centering
 	\begin{subfigure}{1\textwidth}
  		\centering
		\includegraphics[scale=0.30]{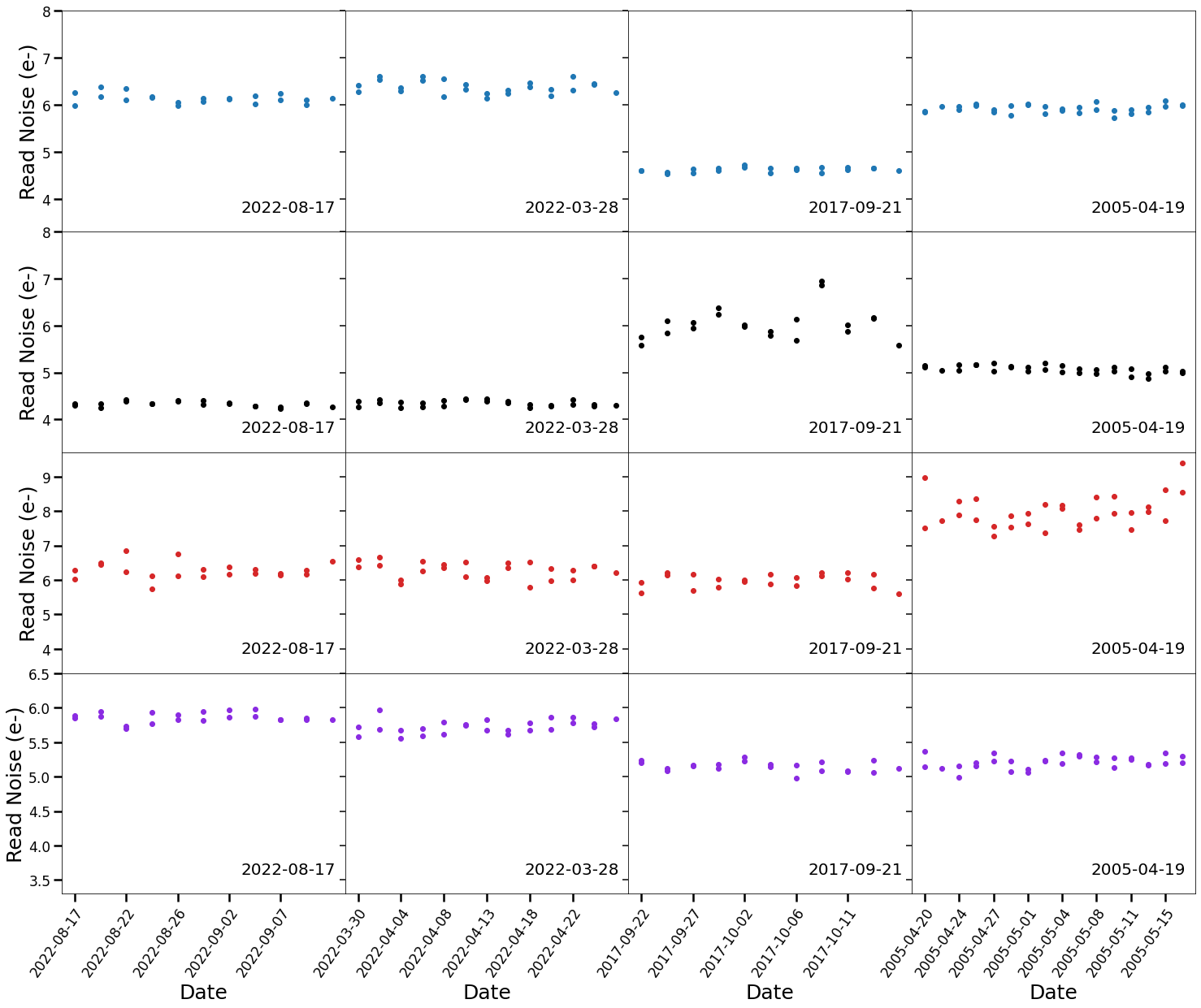}
		\end{subfigure}
    \caption{Read noise plotted over an anneal period for all amplifiers. The first row (blue) shows amplifier A, where column 873 is plotted. The second row (black) is amplifier B where column 1565 is plotted. The third row (red) is amplifier C, where column 373 is plotted and the fourth row (purple) is amplifier D where column 1425 is plotted. Each data point represents the read noise value calculated from an individual bias frame.}
    \label{fig:rntime}
\end{figure}

The read noise for each individual bias from over the course of an anneal period, for each anneal period, is plotted in Figure \ref{fig:rntime}. When looking at column 873 for amplifier A, plotted in the first row, the data indicates that the read noise does not seem to be changing over the time span of an anneal period. For amplifier B, column 1565 exhibited slightly erratic behavior in the 2017 anneal period,  with stable read noise patterns in the pre-SM4 and 2022 anneal periods. Both column 373 for amplifier C and column 1425 for amplifier D, did not show any obvious trends in the read noise values over time. Overall, the data indicates that there is no trend associated with read noise values for particular columns in individual bias frames over an anneal period.

\subsection{Linear Read Noise Decrease in Initial Columns per Amplifier}
The initial read noise vs column number analysis revealed a possible slight linear decrease within the first several hundred science columns for some of the amplifiers. To further investigate this trend, we first trimmed the raw bias data to exclude the physical pre-scans columns, which have lower read noise averages than the science columns. A piecewise linear fit was then conducted on the trimmed data to calculate if and where there is a change in the slope throughout the data set. The piecewise linear fit creates a line of best fit for the data before the break column and another one for the rest of the data. It then returns the break column number and the slope of each line. The Break Column Number is the estimated column number where the break between the first linear decrease is and where the rest of the data continues. Slope 1 refers to the slope of the line with the suspected linear decrease in the first several hundred columns, while Slope 2 refers to the slope of the line that corresponds to the rest of the column data. The initial guess for where the line breaks on each anneal period was approximated using the initial plots.
\subsubsection{Results}

The results of the piecewise linear fit performed for each amplifier in every anneal period are shown in Tables \ref{tab:titleA}, \ref{tab:titleB}, \ref{tab:titleC} and \ref{tab:titleD}.

\begin{figure}[H]
  	\centering
 	\begin{subfigure}{1\textwidth}
  		\centering
		\includegraphics[scale=0.30]{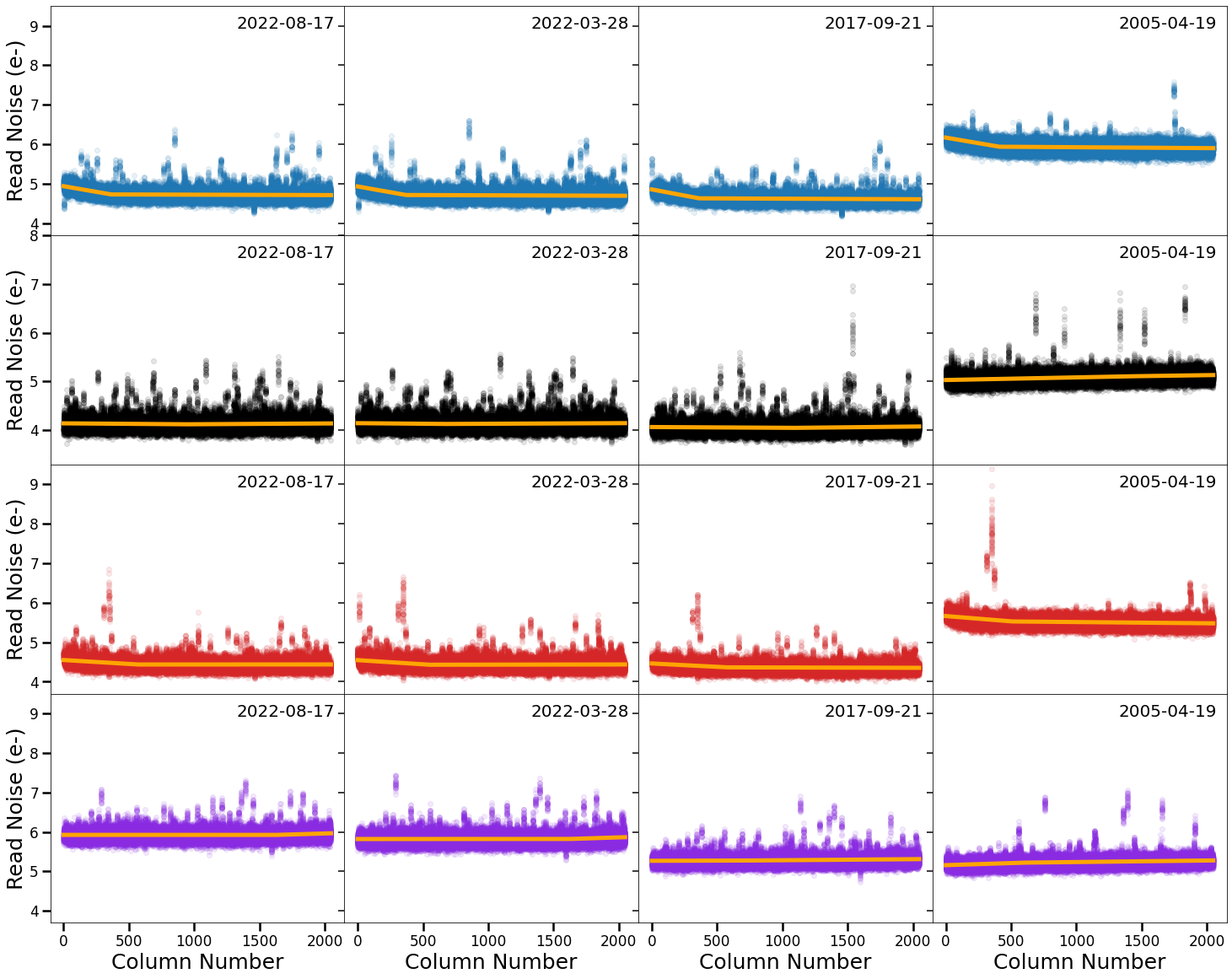}
		\end{subfigure}
    \caption{Read noise as a function of column number for each amplifier, with the orange piecewise linear fit over-plotted.
     The amplifiers are color coded, starting with A as blue, B as black, C as red and D as purple.}
    \label{fig:linefit}
\end{figure}

For all anneal periods, amplifier A had slopes 1 and 2 very close to zero, although Slope 1 exhibited a greater decrease than Slope 2. This indicates that there is in fact a slight linear decrease in the first several hundred columns for amplifier A, breaking between columns 350-450 for a given anneal period. This trend is also seen in amplifier C where slope 2 is even smaller in this case and slope 1 is within the same magnitude as that of amplifier A. These values for amplifier C also indicate a slight linear decrease in the first several hundred columns, but typically breaking between columns 500-600.

\begin{table}[ht]
 \centering
  \begin{threeparttable}
  \captionof{table}{Amplifier A: Piecewise Linear Fit} \label{tab:titleA} 
    \begin{tabular}{| c | c | c | c |}
    	\hline
	Anneal & Break Column Number & Slope 1\tnote{*} & Slope 2\tnote{*} \\ \hline
	2022-08-17 & 362  & -5.75e-04 & -1.026e-05\\
	\hline
 	2022-03-28 & 367 & -5.88e-04  & -1.137e-05\\ 
	 \hline
	 2017-09-21 & 360 & -6.25e-04 & -1.417e-05\\ 
 	\hline
 	2005-04-19 & 407 & -5.654e-04 & -2.22e-05 \\ 
	 \hline
	\end{tabular}
	\begin{tablenotes}
	\footnotesize
	 \item[*] The units for the slope are in $(e^ - / $column number).
	 \end{tablenotes}
	\end{threeparttable}
\end{table}

\begin{table}[ht]
 \centering
  \begin{threeparttable}
  \captionof{table}{Amplifier B: Piecewise Linear Fit} \label{tab:titleB} 
    \begin{tabular}{| c | c | c | c |}
    	 \hline
	 Anneal  & Break Column Number  & Slope 1\tnote{*} & Slope 2\tnote{*} \\ 
	 \hline
 	2022-08-17 & 901 & -2.21e-05  & 1.58e-05\\ 
 	\hline
 	2022-03-28 & 628 & -2.62e-05 & 1.21e-05\\ 
 	\hline
 	2017-09-21 & 1057 & -1.68e-05  & 2.90e-05\\ 
 	\hline
 	2005-04-19 & 1333 &  5.52e-05 & 3.90e-05 \\ 
 	\hline
	\end{tabular}
	\begin{tablenotes}
	\footnotesize
	 \item[*] The units for the slope are in $(e^ - / $column number).
	 \end{tablenotes}
	\end{threeparttable}
\end{table}

\begin{table}[H]
 \centering
  \begin{threeparttable}
  \captionof{table}{Amplifier C: Piecewise Linear Fit} \label{tab:titleC}
    \begin{tabular}{| c | c | c | c |}
    	 \hline
	 Anneal  & Break Column Number  & Slope 1\tnote{*}  & Slope 2\tnote{*} \\ 
 	\hline
 	2022-08-17 & 560 & -2.01e-04  & 9.47e-07\\ 
 	\hline
 	2022-03-28 & 543 & -2.15e-04 & 4.93e-06\\ 
 	\hline
 	2017-09-21 & 564 & -1.74e-04  & -7.89e-06\\ 
 	\hline
	 2005-04-19 & 500 &  -2.72e-04 & -3.09e-05 \\ 
	 \hline
	\end{tabular}
	\begin{tablenotes}
	\footnotesize
	 \item[*] The units for the slope are in $(e^ - / $column number).
	 \end{tablenotes}
	\end{threeparttable}
\end{table}

\begin{table}[H]
 \centering
  \begin{threeparttable}
  \captionof{table}{Amplifier D: Piecewise Linear Fit} \label{tab:titleD} 
    \begin{tabular}{| c | c | c | c |}
 	\hline
 	Anneal  & Break Column Number  & Slope 1\tnote{*} & Slope 2\tnote{*}\\ 
 	\hline
 	2022-08-17 & 1608 & -1.78e-06 & 9.76e-05\\ 
 	\hline
 	2022-03-28 & 1609 & 2.84e-07 & 1.05e-04\\ 
 	\hline
 	2017-09-21 & 636 & 6.87e-06  & 2.69e-05\\ 
 	\hline
 	2005-04-19 & 571 &  1.08e-04 & 3.89e-05 \\ 
 	\hline
	\end{tabular}
	\begin{tablenotes}
	\footnotesize
	 \item[*] The units for the slope are in $(e^ - / $column number).
	 \end{tablenotes}
	\end{threeparttable}
\end{table}

Combining Figure \ref{fig:linefit} and the slope results of the piecewise linear fit in Tables \ref{tab:titleA}, \ref{tab:titleB}, \ref{tab:titleC}, and \ref{tab:titleD}, we determined that amplifiers B and D do not show any linear decrease or increase within the science columns and it is unknown why this linear decrease only appears in amplifiers A and C.

\subsection{Masking Pixels in Columns with Elevated Read Noise}

For this part of the analysis we used the calibration dark and bias frames, as well as superdark and superbias reference files, specifically the DQ and SCI arrays, to better understand the elevated read noise values that emerged for particular columns over the various anneal periods. In the 2022-08-17 anneal period, the column with the highest read noise values (average of 6.2 e$^-$) for amplifier A was column 873. This column was thus selected for further analysis while we attempt to reduce the read noise in each raw bias frame using various methods.

\subsubsection{Masking Hot Pixels in Dark Frames}
The first strategy was to find hot pixel trails, mask these pixels and then plot the resulting read noise values for column 873. Hot pixel trails are two or more pixels being affected by dark current bleeding into other pixels in the direction it is read out. Superdark DQ arrays are useful for finding these trails, however since superdarks are trimmed to exclude physical pre-scans and virtual overscans, it was necessary for us to trim our bias frames too for easy comparison of the column being analyzed. When investigating column 873 on the raw bias image, to get the corresponding column in the superdark frame, we needed to subtract 24 columns corresponding to the physical pre-scans. Therefore, column 873 in the bias frame corresponds to column 849 in the superdark frame. 

Two hot pixel trails were found for column 873 in the 2022-08-17 anneal period, pixel rows 397-400 and 1571-1575. These hot pixel trails were also found in the 2022-03-28 anneal period, but were not seen on the 2017 or pre-SM4 anneal periods, potentially explaining why there is an increase in read noise for the recent anneal periods.

\subsubsection{Masking Unstable Hot Columns in Bias Frames}
The second strategy to decrease the read noise in elevated columns was to find unstable hot columns, flagged as 128 in the superbias DQ array, and mask the portion of the column that is flagged as unstable to see if this would have any effect on the elevated read noise. We decided to start with the 2022-03-28 anneal period. Using a side-by-side comparison of the superbias SCI array and the superbias DQ array in DS9, the unstable hot column was visible and ranged approximately from row zero to row 1590. In the DQ array, these specific row values were flagged as an unstable hot column. To remove this column we decided to increase the range of rows that were masked, up to 1611, to not miss any pixels of the unstable pixel trail. The same rows were masked for column 873 on the 2022-08-17 anneal period. 

\subsubsection{Results}

\begin{figure}[H]
  	\centering
 	\begin{subfigure}{1\textwidth}
  		\centering
		\includegraphics[scale=0.5]{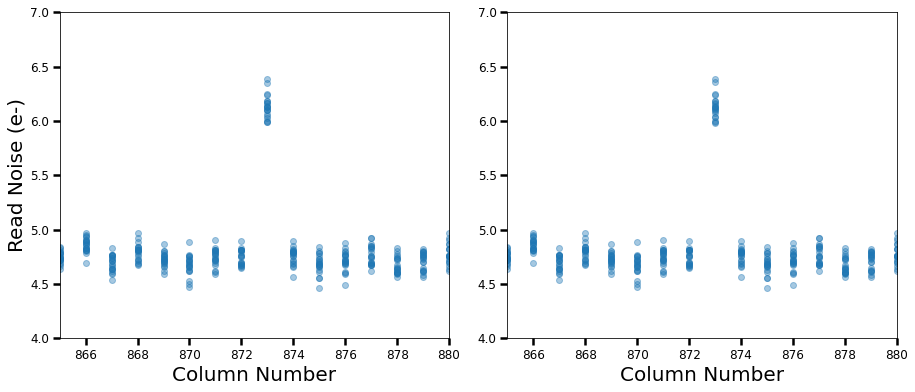}
		\end{subfigure}
    \caption{Read noise values before and after masking hot pixel trails in column 873, amplifier A, from the 2022-08-17 anneal period. This is a zoomed in version of the data, focusing on the elevated read noise values seen in column 873, as well as the read noise levels of the surrounding columns. The first plot (left) is the original data, before the column was masked. The second plot (right) is after the two hot pixel trails were masked, where little to no change is seen.}
    \label{fig:rnbeforeafter}
\end{figure}

The initial method used to reduce the read noise in column 873 was not successful. There was little to no change in the read noise values when the two hot pixel trails identified were masked. Therefore, we can deduce that hot pixel trails found in the superdark are not the primary reason for elevated read noise levels in certain columns. Figure \ref{fig:rnbeforeafter} shows the comparison of the read noise values for column 873 before and after masking the trails. 

\begin{figure}[H]
  	\centering
 	\begin{subfigure}{1\textwidth}
  		\centering
		\includegraphics[scale=0.5]{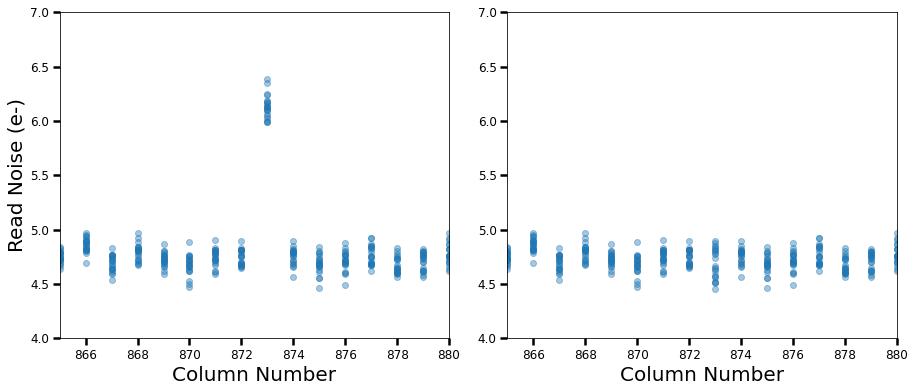}
		\end{subfigure}
    \caption{Read noise values before and after masking rows 0 to 1611 in column 873 for amplifier A, from the 2022-08-17 anneal period. The first plot (left) is the original data, before the column was masked. The second plot (right) shows the reduced noise after the rows were masked.}
    \label{fig:873august}
\end{figure}

\begin{figure}[H]
  	\centering
 	\begin{subfigure}{1\textwidth}
  		\centering
		\includegraphics[scale=0.45]{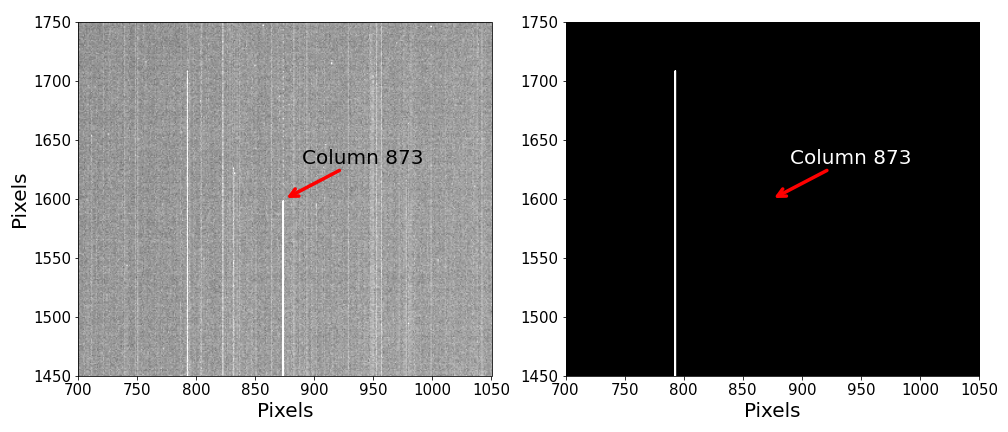}
		\end{subfigure}
    \caption{Images of the superbias SCI frame (left) and the superbias DQ frame (right) for the 2022-08-17 anneal period. The red arrows are pointing to column 873 on both frames. In this case, the hot column seen in the SCI frame is not flagged as unstable in the DQ frame.}
    \label{fig:202208superbias}
\end{figure}

The second method reduced the read noise values for column 873 to average read noise values as seen in Figure \ref{fig:873august}. Looking at the superbias SCI and DQ frames for the 2022-08-17 anneal period, Figure \ref{fig:202208superbias}, the hot column is seen in the SCI frame, however this column for this anneal period was not flagged as an unstable hot column (flag 128) in the DQ array. 

\begin{figure}[H]
  	\centering
 	\begin{subfigure}{1\textwidth}
  		\centering
		\includegraphics[scale=0.5]{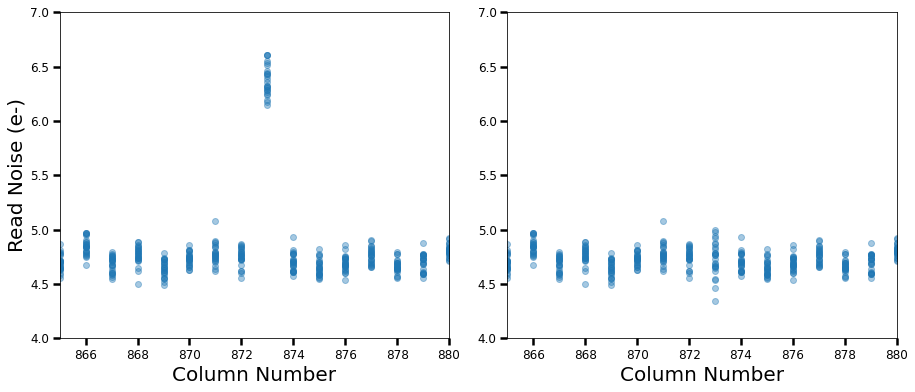}
		\end{subfigure}
    \caption{Read noise values before and after masking rows 0 to 1611 in column 873 for amplifier A, from the 2022-03-28 anneal period. The first plot (left) is the original data, before the column was masked. The second plot (right) shows the reduced noise after the rows were masked.}
    \label{fig:873march}
\end{figure}

\begin{figure}[H]
  	\centering
 	\begin{subfigure}{1\textwidth}
  		\centering
		\includegraphics[scale=0.45]{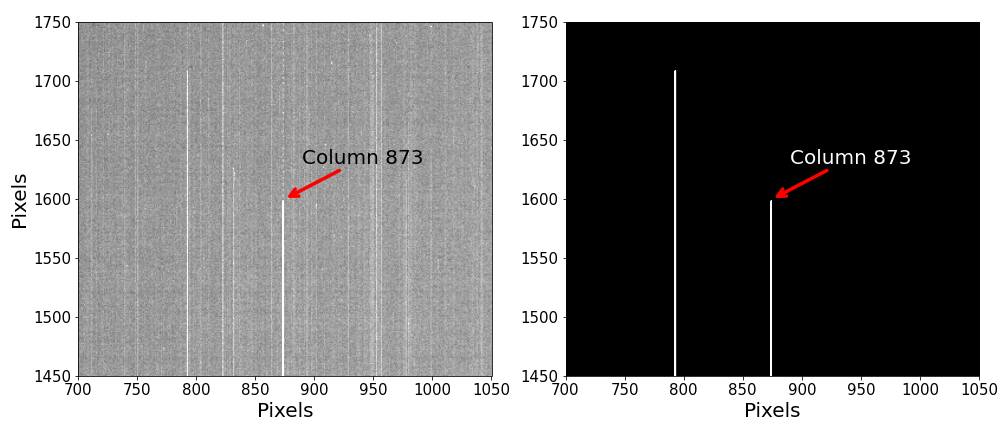}
		\end{subfigure}
    \caption{Images of the superbias SCI frame (left) and the superbias DQ frame (right) for the 2022-03-28 anneal period. The red arrows are pointing to column 873 on both frames. In this case, the hot column seen in the SCI frame is flagged as an unstable column in the DQ frame.}
    \label{fig:202203superbias}
\end{figure}

The same steps for the second method were taken to reduce the read noise values of column 873 for the 2022-03-28 anneal period as seen in Figure \ref{fig:873march}. However, when we analyzed the SCI and DQ superbias frames for this anneal period, we noticed the same hot trail in the SCI frame for column 873, but this time it was flagged as an unstable hot column in the DQ frame (Figure \ref{fig:202203superbias}). The flagging of unstable hot columns depends on the stability of the hot columns over an anneal period.

Although we find that there is no dependence of read noise with column number, these elevated columns could introduce outliers in the read noise calculation with respect to each amplifier, and should therefore be masked.

\section{Conclusion} \label{s:conclusion}
In this study we analyzed the read noise of each amplifier throughout four anneal periods. We determined that read noise values with respect to each amplifier do not depend on column number. This is due to the direction the pixels are read out and therefore the read noise only depends on row number and increases across the detector. When we examined the read noise vs column number plots generated for each amplifier, we noticed that the pre-scans have a lower average read noise values than the rest of the science image for all anneal periods. We concluded that because the pre-scans have no dark current accumulated, there would be less noise present in this part of the WFC CCD. Therefore, we suggest not including the physical pre-scan columns when calculating the read noise to get more accurate results during calibration.

We decided to expand and explore other characteristics of the data such as how the read noise of a column evolves over an anneal period. We chose four different columns, one for each amplifier, that have different characteristics. For amplifier A we chose column 873 which was elevated in both 2022 anneal periods. Column 1565 was chosen for amplifier B and was only elevated in the 2017-09-21 anneal period. For amplifier C we worked with column 373 which was elevated in all anneal periods observed. Lastly, column 1425 was chosen for amplifier D which was not elevated. The data revealed no trend between the read noise of a column throughout an anneal period for any of the chosen columns.  

Looking at the generated plots of read noise vs column number, the data showed a slight decrease of the read noise values for amplifiers A and C. We investigated this further by trimming the physical pre-scans and using a piecewise linear function to get a line of best fit for the data of each amplifier throughout the anneal periods. The piecewise linear function allowed us to calculate the break column number of when the data changed from decreasing values to steady values, and also the slopes before and after the break column number. We concluded that both amplifiers A and C have a small linear decrease in the first several hundred columns, but we are unsure what is causing this behavior that is not seen in either amplifier B or D.

Lastly, there were columns in the original data that had elevated read noise values in all the bias files. To better understand what causes the high read noise values, we decided to try two different methods to mask certain pixels within the column and see if it would reduce the elevated read noise values. We worked with column 873 on amplifier A since it had elevated read noise in both 2022 anneal periods. The first method consisted of masking trails found in the superdark DQ array. However, this method was unsuccessful at reducing the read noise of the column because very few pixels were masked. For the second method we looked at the superbias SCI array where we found unstable hot pixels flagged in column 873. We masked the portion of the unstable hot column and this lowered the read noise to average values. We checked the superbias DQ arrays to see if this hot pixel trail was flagged as an unstable hot column and found that in the 2022-08-17 anneal period it was not flagged as an unstable hot column, but in the 2022-03-28 anneal period it was. This happens because the flagging of unstable hot columns depends on the stability of the hot columns over an anneal period. From these results we conclude that unstable hot columns should be masked to reduce the read noise values for calibration and the creation of reference files. 

\section{Acknowledgements}

We thank Norman Grogin and Roberto Avila for assistance, feedback and support throughout the project. We also thank Jenna Ryon, David Stark, Ralph Bohlin and Gagandeep Anand for providing comments to improve the report.

For this report we used the following: \texttt{jupyter} \citep{jupyter}, \texttt{numpy}  \citep{cite_numpy}, \texttt{pandas} \citep{pandas}, \texttt{astropy} \citep{astropy}, and  \texttt{matplotlib} \citep{matplotlib}.

\bibliographystyle{apj}
\bibliography{guzman_readnoise}

\end{document}